\documentstyle[preprint,aps,prc]{revtex}

\newcommand{\be}{\begin{equation}}
\newcommand{\ee}{\end{equation}}

\newcommand{\ia}{^{\mbox{}}_{\alpha}}
\newcommand{\ib}{^{\mbox{}}_{\beta}}
\newcommand{\iaa}{^{\mbox{}}_{\alpha\alpha}}

\begin{document}

\draft

\title{An HFB scheme in natural orbitals}

\author{P.--G. Reinhard$^{a}$, M. Bender$^{b}$, 
        K. Rutz$^{b}$, and J. A. Maruhn$^{b}$}

\address{${}^{a}$ Institut f\"ur Theoretische Physik, 
         Universit\"at Erlangen\\
         Staudtstr.\ 7, D-91058 Erlangen, Germany}

\address{${}^{b}$ Institut f\"ur Theoretische Physik, 
         Universit\"at Frankfurt\\
         Robert-Mayer-Str. 10, D-60325 Frankfurt, Germany.}

\date{14 April 1997}

\maketitle

\begin{abstract}
We present a formulation of the Hartree-Fock-Bogoliubov (HFB) equations
which solves the problem directly in the basis of natural orbitals. This
provides a very efficient scheme which is particularly suited for large
scale calculations on coordinate-space grids.
\end{abstract}

The production of new nuclei towards the drip lines and in the
super-heavy region (for a review see \cite{revexotic}) has raised a
growing interest in the refinement of nuclear mean-field models to
accommodate the larger body of experimental data. At the level of
precision reached nowadays, a correct treatment of pairing becomes a
crucial ingredient \cite{revpair}. This calls for a full variational
optimization of the pairing wave function, what is known as the
Hartree-Fock-Bogoliubov (HFB) treatment \cite{RinSch,Doba}. It is the
aim of this short note to present an efficient solution scheme for the
HFB equations which relies on a direct variational optimization within
the natural orbital basis. The following discussion is formulated for
one sort of Fermions. Nuclear applications will use that scheme for
neutrons and protons separately. A generalization to proton-neutron
pairing is obvious.

Starting point is the BCS ansatz for the wave function of a pairing
many-body system
\be
  |\Phi\rangle
  =
  \prod \left(u\ia+v\ia a^+_\alpha a^+_{\bar{\alpha}}\right)
       |{\rm Vac}\rangle
\label{BCSansatz}
\ee
where $a^+_\alpha$ generates a particle in the state $\varphi_\alpha$,
$\bar{\alpha}$ is the time reversed partner of $\alpha$,
$v\ia$ the occupation amplitude for the state, and
$u\ia=\sqrt{1-v_\alpha^2}$ the non-occupation
amplitude. The ansatz (\ref{BCSansatz}) requires that the 
single-particle states are orthonormalized,
\be
  \left(\varphi\ia|\varphi\ib\right)
  =
  \delta_{\alpha\beta}^{\mbox{}}
  \quad,
\label{orthon}
\ee
and that the occupations add up to the total particle number
$\sum v_\alpha^2=N$.
These two presuppositions will have to be added as boundary conditions
in the variation later on. Note, furthermore,
that we have assumed here the case of
a time-even state (ground state of even-even nuclei) for which $u\ia$
and $v\ia$ can be chosen real and for which we can construct the
time-reversed wavefunction as
$\varphi_{\bar{\alpha}}=-i\hat{\sigma}_y\varphi^*_\alpha$
where $\hat{\sigma}_y$ is a Pauli spin matrix.

The BCS ansatz (\ref{BCSansatz}) carries only one-body information
which is summarized in the two key densities: the one-body density
operator
\be
  \hat{\rho}
  =
  \sum v_\alpha^2 \varphi\ia\varphi^+_\alpha
  \quad,
\label{onedens}
\ee
and the pair-density operator
$\hat{\chi}=\sum_\alpha u\ia v\ia \varphi\ia\varphi^+_\alpha$
from which we are going to use only the local part
\be
  \chi(r)
  =
  \sum u\ia v\ia \varphi^{+}_{\alpha}(r)\varphi\ia(r)
  \quad.
\label{pairdens}
\ee
Thereby, we have employed time-reversal symmetry
which also yields the property $\chi^*(r)=\chi(r)$. A
discussion of the physical content of the pair density can be
found in \cite{newdoba}.
Several useful properties are further discussed
in \cite{Doba}. Note that the one-body density matrix
(\ref{onedens}) is diagonal in the chosen representation which means
that we are dealing with the natural orbital basis. It is noteworthy
that the pair density is also diagonal in the same basis. This is
guaranteed by the relation $[\hat{\rho},\hat{\chi}]=0$ which can be
derived on general grounds \cite{Doba}.

The standard solution scheme for the HFB problem deals with a four times
as large super-density-matrix which encompasses $\hat{\rho}$ as well
as $\hat{\chi}$, and it proceeds in three subsequent unitary
transformations \cite{RinSch}. There is a treatment in terms of wave
functions which is particularly suited for coordinate space
representations and which uses a double set of single-particle
wavefunctions, one for the occupation part and one for the
non-occupation part \cite{Doba}. Both schemes are plagued by the
doubling of the representation which adds substantial overload to the
calculations and which becomes very cumbersome in deformed nuclei.
The BCS-ansatz (\ref{BCSansatz}) can be
formulated in terms of a single set of wave functions, the natural
orbitals, and a few occupation amplitudes. It is possible to keep the
scheme at that level of expense by a direct variational exploitation of
the ansatz, as will be shown in the following.

For simplicity, we work here with the case that the energy 
separates into a mean-field part and a pairing part as
\be
  E
  =
  E_{\rm mf}(\hat{\rho}) + E_{\rm pair}(\hat{\chi})
  \quad.
\ee
Although the present scheme is applicable under more general conditions,
we use now a particular model for the pairing energy, a
volume pairing with a zero-range force
\be
  E_{\rm pair}
  =
  \frac{1}{4}V_{\rm P}\int d^3r \, \chi(r)^2
\label{Epair}
\ee
which employs only the local pair density (\ref{pairdens}). The energy
functional determines the mean-field Hamiltonian
and pair potential as the functional derivatives
\be
  \hat{h}_{\rm mf}
  =
  \frac{\partial E}{\partial\hat{\rho}}
  \quad,\quad
  \Delta(r)
  =
  \frac{\partial E}{\partial\chi(r)}
  =
  \frac{1}{2}V_{\rm P} \chi(r)
  \quad.
\ee
Note that we obtain a local pair potential for the particular pairing
functional (\ref{Epair}).

The variation with respect to the occupation amplitudes $v\ia$ yields%
$$
  0
  =
  \frac{\partial (E-\epsilon_{\rm F}N)}{\partial v\ia}
  =
  4 v\ia (h\iaa - \epsilon_{\rm F})
  + 2 \left(\frac{v_\alpha^2}{u\ia}-u\ia\right) \Delta\iaa
$$
where $\epsilon_F$ is the Lagrange multiplier for the particle number
constraint.
This equation can be resolved in the standard manner and yields
\be
  \left\{\begin{array}{c} v\ia \\ u\ia \end{array}\right\}
  =
  \sqrt{\frac{1}{2}\mp \frac{1}{2}
      \frac{h\iaa-\epsilon_{\rm F}}
           {\sqrt{(h\iaa-\epsilon_{\rm F})^2+\Delta^2_{\alpha\alpha}} } }
   \quad.
\label{uveq}
\ee
Note that the gap potential $\Delta$ does not necessarily commute with
the  mean-field Hamiltonian $\hat{h}$. Only the diagonal elements in the
natural orbital basis enter and no information about the non-diagonal
elements is needed.

The variation with respect to the single particle wavefunctions needs
to take care of the orthonormality (\ref{orthon}) which is done by
adding $-\sum_{\alpha\beta}\lambda^{\mbox{}}_{\alpha\beta}
\left(\varphi\ia|\varphi\ib\right)$.  The thus constrained variation
yields
\be
  \hat{\cal H}\ia \varphi\ia
  =
  \textstyle{\sum_\beta}
  \lambda^{\mbox{}}_{\alpha\beta} \varphi\ib
\label{cmfeq}
\ee
with a generalized mean-field Hamiltonian
\be
  \hat{\cal H}\ia \varphi_\alpha
  =
  \frac{\partial E}{\partial \varphi^+_\alpha}
  =
  \Big[ v^2_\alpha \hat{h}_{\rm mf} + u\ia v\ia \Delta(r)
  \Big] \varphi_\alpha
  \quad.
\label{genmf}
\ee
This
$\hat{\cal H}\ia$ is a state dependent Hamiltonian and the full matrix
of Lagrangian multipliers needs to be taken into account.
Thereby it is crucial that they
constitute a symmetric matrix
$\lambda^{\mbox{}}_{\alpha\beta} = \lambda^{\mbox{}}_{\beta\alpha}$.
This allows to symmetrize explicitely
\be
  \lambda^{\mbox{}}_{\alpha\beta}
  =
  \frac{1}{2}
  \left(\varphi\ib|\hat{\cal H}\ia+\hat{\cal H}\ib|\varphi\ia\right)
\label{avlambda}
\ee
which links pairwise all $\alpha$ with $\beta$ and thus 
eq.~(\ref{cmfeq}) delivers a decisive problem.

Altogether, the equations (\ref{uveq}), (\ref{genmf}), (\ref{cmfeq}), and
(\ref{avlambda}) constitute the HFB equations formulated directly in the
natural orbital basis. This nonlinear problem is best
solved iteratively. We propose an interlaced iteration
employing the efficient damped gradient step which is best suited for
coordinate space techniques \cite{dampgrad}:
\begin{enumerate}
 \item\label{step1}
  Start from a given set $\{\varphi\ia\,v\ia\}$.
 \item
  Compute the densities $\hat{\rho}$ and $\chi(r)$, and subsequently the
  corresponding Hamiltonians $\hat{h}_{\rm mf}$ and $\Delta(r)$.
 \item
  Compute the new $v\ia$ and $u\ia$ according to to Eq.~(\ref{uveq}).
 \item
  Compute the action of the state-dependent mean field $\hat{\cal H}\ia$
  and store the resulting set
  $\{\hat{\cal H}\ia\varphi\ia\}$.
 \item
  Compute the matrix of constraints $\lambda^{\mbox{}}_{\alpha\beta}$,
  Eq.~(\ref{avlambda}).
 \item
  Perform the damped gradient step with
  \be
    \varphi\ia
    \leftarrow
    {\cal O}\{ \varphi\ia  -
    {\cal D}_\alpha[\hat{\cal H}\ia\varphi\ia
       - \textstyle{\sum_\beta}\,\,
        \lambda^{\mbox{}}_{\alpha\beta}\varphi\ib]
            \}
  \ee
  where ${\cal O}$ means orthonormalization
  and ${\cal D}_\alpha$ is an appropriate
  damping operator.
  This completes the scheme and returns to the starting point,
  step~\ref{step1}.
\end{enumerate}
The state-dependent Hamiltonian (\ref{genmf}) requires a state-dependent
damping for which we generalize the form of \cite{dampgrad} to%
\be
 {\cal D}_\alpha
 =
 \frac{x^{\mbox{}}_0}
 {v_\alpha^{2}(50\,{\rm MeV}+\hat{T})
  +u_\alpha v_\alpha\frac{1}{2}{\rm max}\{\Delta(r)\}}
\ee
where $\hat{T}$ is the kinetic energy operator and $x_0\!\approx\!0.2$ a
numerical parameter. With that choice, we have implemented this scheme
successfully into a spherical Skyrme-Hartree-Fock code and tested it
extensively for a variety of Skyrme forces and nuclei from
$^{16}{\rm O}$ to the isotopes of Pb, including proton rich as well as
neutron rich exotic nuclei. The scheme proves to be reliable and very
efficient. It allows a fast computation of the HFB ground state, costing
only 20\% more iterations than the much simpler BCS approximation, and
each iteration as such is as fast as in the BCS case because only one
set of single particle wavefunctions is handled. We thus have a
promising alternative HFB scheme which can simplify large scale
calculations of deformed nuclei.

\end{document}